\documentclass[fleqn,usenatbib]{mnras}

\usepackage{newtxtext,newtxmath}

\usepackage[T1]{fontenc}

\DeclareRobustCommand{\VAN}[3]{#2}
\let\VANthebibliography\thebibliography
\def\thebibliography{\DeclareRobustCommand{\VAN}[3]{##3}\VANthebibliography}

\usepackage{graphicx}	
\usepackage{amsmath}	
\usepackage{xcolor} 
\usepackage{caption}
\usepackage{subcaption}

\newcommand{\qstokes}{\ensuremath{\mathcal{Q}}}
\newcommand{\ustokes}{\ensuremath{\mathcal{U}}}

\title[Flares in the Galactic centre II]{Flares in the Galactic centre II: polarisation signatures of flares at mm-wavelengths}

\author[M. Najafi-Ziyazi et al.]{
Mahdi Najafi-Ziyazi$^{1}$\thanks{mahdi.najafiziyazi@student.uva.nl},
Jordy Davelaar$^{2,3}$,
Yosuke Mizuno$^{4,5,6}$,
and Oliver Porth$^{1}$\thanks{Corresponding Author. o.porth@uva.nl}
\\
$^{1}$ Anton Pannekoek Institute for Astronomy, University of Amsterdam, Science Park 904, 1098 XH, Amsterdam, The Netherlands\\
$^{2}$Center for Computational Astrophysics, Flatiron Institute, 162 Fifth Avenue, New York, NY 10010, USA\\
$^{3}$Department of Astronomy and Columbia Astrophysics Laboratory, Columbia University, 550 W 120th St, New York, NY 10027, USA\\
$^{4}$Tsung-Dao Lee Institute, Shanghai Jiao-Tong University, Shanghai, 520 Shengrong Road, 201210, People's Republic of China\\
$^{5}$School of Physics \& Astronomy, Shanghai Jiao-Tong University, Shanghai, 800 Dongchuan Road, 200240, People's Republic of China\\
$^{6}$Institut f\"{u}r Theoretische Physik, Goethe Universit\"{a}t, Max-von-Laue-Str. 1, 60438 Frankfurt am Main, Germany\\
}

\date{Accepted XXX. Received YYY; in original form ZZZ}

\pubyear{2023}

\begin{document}
\label{firstpage}
\pagerange{\pageref{firstpage}--\pageref{lastpage}}
\maketitle

\begin{abstract}
Recent polarimetric mm-observations of the galactic centre by \cite{wielgus22} showed sinusoidal loops in the $\mathcal{Q-U}$ plane with a duration of one hour. The loops coincide with a quasi-simultaneous X-ray flare.  
A promising mechanism to explain the flaring events are magnetic flux eruptions in magnetically arrested accretion flows (MAD).  
In our previous work \citep{2021MNRAS.502.2023P}, we studied the accretion flow dynamics during flux eruptions.  Here, we extend our previous study by investigating whether polarization loops can be a signature produced by magnetic flux eruptions. \\
We find that loops in the $\mathcal{Q-U}$ plane are robustly produced in MAD models as they lead to enhanced emissivity of compressed disk material due to orbiting flux bundles. 
A timing analysis of the synthetic polarized lightcurves demonstrate a polarized excess variability at timescales of $\simeq 1~\rm hr$.  The polarization loops are also clearly imprinted on the cross-correlation of the Stokes parameters which allows to extract a typical periodicity of $30~\rm min$ to $1~\rm hr$ with some evidence for a spin dependence.  These results are intrinsic to the MAD state and should thus hold for a wide range of astrophysical objects.  
A subset of GRMHD simulations without saturated magnetic flux (single temperature SANE models) also produces $\mathcal{Q-U}$ loops. However, in disagreement with the findings of \cite{wielgus22}, loops in these simulations are quasi-continuous with a low polarization excess.  
\end{abstract}


\section{Introduction}

The centre of our galaxy harbours one of the brightest radio sources in our sky, which is thought to coincide with the location of a putative supermassive black hole called Sagittarius A* (Sgr A*). The source has been extensively studied since its discovery in the 1970s \citep{Balick1974}. More recently, the Event Horizon Telescope Collaboration reconstructed the first direct image of Sgr A* \citep{EventHorizonTelescopeCollaborationAkiyamaEtAl2022,EventHorizonTelescopeCollaborationAkiyamaEtAl2022d}. The image shows a ring-like structure with a darkening in its centre, interpreted as gravitational lensing of the black hole's event horizon -- an effect often referred to as the black hole's 'shadow' \citep{Bardeen1973,Luminet1979,Falcke2000}. The size of the ring is $50 ~\mu$as, which agrees within 10 per cent with the prediction for a $4.1\times10^6 M_\odot$ Kerr black hole whereby the mass is well constrained through the monitoring of stellar orbits \citep{2019Sci...365..664D,CollaborationAbuterEtAl2019}. 

Sgr A* exhibits variability on all frequencies at timescales from minutes to hours \citep[e.g.][]{MarroneBaganoffEtAl2008,WitzelMartinez2018,HaggardNynkaEtAl2019,IwataOkaEtAl2020,EventHorizonTelescopeCollaborationAkiyamaEtAl2022c,WielgusMarchiliEtAl2022}. The strongest variable signatures are manifested as X-ray flares whereby the flux has been seen to increase several hundred-fold over a timescale of $\sim 20$ minutes \citep{HaggardNynkaEtAl2019}. 
The GRAVITY experiment observed Sgr A* during NIR flares and found a bright component orbiting clockwise around the compact object at approximately ten gravitational radii \cite{GravityCollaborationAbuterEtAl2018,TheGRAVITYCollaborationAbuterEtAl2023}. Besides centroid motion, they also measured the polarization during the flare, which shows periodic variation in Stokes $\mathcal{Q}$ and $\mathcal{U}$, indicative of a hotspot that rotates through a poloidal magnetic field. 

Similar $\mathcal{Q-U}$ loops have been found at mm-wavebands \citep{Marrone2006,wielgus22}. The most recent work by \cite{wielgus22} analysed the Atacama Large Millimeter Array (ALMA) lightcurve obtained during the EHT campaign in April 2017 (see also \citet{WielgusMarchiliEtAl2022}). The data of April 11th commences just after an X-ray flare and indicates a drop of mm flux and a sinusoidal pattern in both Stokes $\mathcal{Q}$ and $\mathcal{U}$, leading to clockwise motion in a $\mathcal{Q-U}$ diagram (in the following: ``the ALMA loop'').
In \cite{VosMoscibrodzkaEtAl2022}, the authors argue that the observation can be explained by a hotspot orbiting the black hole through a poloidal magnetic field. Although this simplified model successfully matches many aspects of the observation, it is uncertain if accretion flows will give rise to similar $\mathcal{Q-U}$ loops if the flow geometry is more complex and dynamically varying. We must rely on general relativistic magnetohydrodynamic (GRMHD) simulations to answer this question. 

Although Sgr A* is one of the brightest radio sources in the sky, its total bolometric luminosity is orders of magnitude lower than its Eddington luminosity such that Sgr A* is in the regime of Radiatively Inefficient and Advection Dominated Accretion Flows (RIAF/ADAF) \citep{Ichimaru1977, Narayan1994}. 
The EHT measurements \citep{EventHorizonTelescopeCollaborationAkiyamaEtAl2022} put constraints on the state of the accretion flow, these favour a magnetically arrested disk (MAD) model with an accretion rate $(5.2-9.5)\times 10^{-9}\rm M_\odot/yr$ \citep{EventHorizonTelescopeCollaborationAkiyamaEtAl2022c}.  
Similarly, the EHT measurements for M87* also favour the MAD state \cite{CollaborationAkiyamaEtAl2019d,Collaboration2021}.  

In a MAD, the magnetic flux on the horizon has reached a maximum for a given accretion rate at which the accretion disk can be magnetically choked \citep{Igumenshchev2003,Tchekhovskoy2011}, in contrast to Standard And Normal Evolution models (SANE). When a MAD model reaches its saturation state, the jet's magnetic pressure and the disk's ram pressure are in quasi-equilibrium. Accretion proceeds through spiral modes and the dynamics are highly intermittent:
as more magnetic flux is accreted onto the black hole, eventually the jet will expel parts of its magnetic flux back into the disk in a bursty flux eruption.  
Whether the flux eruption is governed by interchange instability between plasma on horizon penetrating field lines and the disk \citep[e.g.][]{McKinneyTchekhovskoy2012} or via an equatorial current sheet \citep{RipperdaLiskaEtAl2022} or both has not been settled conclusively.  
In any case, the flux eruption generates an orbiting flux tube which is subsequently dissolved in the accretion flow, most likely by Rayleigh Taylor-type instabilities \citep{RipperdaLiskaEtAl2022,ZhdankinRipperdaEtAl2023}. After the flux eruption, accretion of matter and flux proceeds and the magnetic flux builds up until the next flux eruption event, leading to a repeating cycle of dissipation and accumulation of magnetic flux. The dynamics of orbiting flux tubes were studied in some detail in our first paper of this series \citep{2021MNRAS.502.2023P} and by others \citep[e.g.][]{BegelmanScepiEtAl2022,ChatterjeeNarayan2022c}.  

In contemporary MAD GRMHD simulations, the recurrence time of flux eruptions is $\approx 1000~GM/c^3$ corresponding to $\simeq 5\, 1/2~\rm hours$ for Sgr A*.  
Since near-infrared flares occur on a similar timescale, ie. $2-4$ per day \citep[e.g.][]{GenzelSchodelEtAl2003,Dodds-EdenGillessen2011} -- with roughly one in four also having an X-ray counterpart \citep[e.g.][]{BaganoffBautz2001,HornsteinMatthewsEtAl2007,BoyceHaggardEtAl2019} -- flux eruptions seem a very good candidate for the physical process triggering such flares.  
 Hence there is great interest in magnetic flux eruptions to explain AGN flares and several models have been applied to M87* as well as Sgr A* \citep{DexterTchekhovskoyEtAl2020,2020ApJ...900..100R,ChatterjeeMarkoffEtAl2020a,2021MNRAS.502.2023P,ScepiDexter22,JiaRipperdaEtAl2023,ZhdankinRipperdaEtAl2023}. In principle, the dissipation of magnetic energy can accelerate particles to non-thermal energies giving rise to efficient synchrotron emission up to X-rays. Additionally, the ejected flux tube orbits in the disk, which might account for the centroid motion observed by GRAVITY. While several studies have addressed the behaviour of total intensity at various wavelengths and various prescriptions for the underlying particle distribution, the polarisation signatures of flux eruptions in MAD simulations are not yet well understood.

In this paper, we use GRMHD simulations in the SANE and MAD regime to characterize their linear polarization time-domain signatures with a particular focus on $\mathcal{Q-U}$ loops. In section \ref{sec:methods}, we will describe our GRMHD models and our polarized radiative transfer code. In section \ref{sec:results}, we will describe our results. In section \ref{sec:discussion}, we will discuss our findings and summarize our conclusions.

\section{Methods}\label{sec:methods}

\subsection{General relativistic Magnetohydrodynamic models}

We analyze a number of ideal GRMHD simulations both in the SANE and in MAD regimes carried out with the \texttt{BHAC} code \footnote{https://bhac.science} \citep{PorthOlivares2017,OlivaresPorthEtAl2019}.  We cover different black hole spins $a_*\in\{-0.9375,-0.5,0,0.5,0.9375\}$, where $a_*:={J}/(Mc)$, with $J$ being the black hole angular momentum, $M$ the black hole mass, and $c$ the speed of light. Negative spin indicates counter-alignment between the angular momenta of the accreting matter and black hole as customary. The simulations are part of the EHT simulation library and have previously yielded model comparison for the EHT 2017 Sgr A* campaign \citep{EventHorizonTelescopeCollaborationAkiyamaEtAl2022a}.  
The simulations are performed in spherical horizon penetrating Kerr-Schild coordinates $r\in[1.19,2500]r_{\rm g}$, where $r_{\rm g}:=G M/c^2$, with $G$ being the gravitational constant. We perform our analysis over the time span $t\in [15,30]\times 1000\, r_{\rm g}/c$ corresponding to roughly $3.5\, \rm days$ when scaled to Sgr A*. Through three static mesh refinement levels, we obtain an effective grid resolution of $N_r\times N_\theta \times N_\phi=384\times 192\times 192$ cells which is sufficient to resolve the magneto-rotational instability throughout the evolution \citep{PorthChatterjeeEtAl2019a}. For details of the employed numerical methods and setup, we refer to \cite{PorthOlivares2017,OlivaresPorthEtAl2019,Mizuno2021} and \cite{EventHorizonTelescopeCollaborationAkiyamaEtAl2022a}.

\subsection{General relativistic ray tracing}

The GRMHD simulations provide a dynamical 3D model for the structure of the accretion flow. To compare our models to observations, we post-process the GRMHD simulations with our general relativistic ray-tracing code {\tt RAPTOR} \footnote{https://github.com/jordydavelaar/raptor} to compute synthetic polarized images and light curves. {\tt RAPTOR} computes null geodesics in curved spacetimes from a virtual camera outside our simulation domain. The camera is assigned pixels, and each is given an initial wave vector, after which the geodesic equation is solved to find their corresponding trajectories. Along these paths, we then solve the polarized radiative transport equation to obtain synthetic images for all four Stokes parameters, $\mathcal{I}$, $\mathcal{Q}$, $\mathcal{U}$, and $\mathcal{V}$. 
Observables are produced for a fiducial EHT frequency of $229.1~\rm GHz$ and we consider thermal synchrotron emission (thus e.g. no Bremsstrahlung), absorption, Faraday rotation and conversion. 
The emission and absorption coefficients are obtained from the fit functions given by \cite{Dexter2016} and rotativities from \cite{Shcherbakov2008}.  
In this study, we ignore effects caused by the finite propagation speed of the rays and assume the MHD state remains fixed during raytracing (known as the ``fast light'' approximation).  
Our spacetime is described by a Kerr-Schild metric for a rotating black hole. We set our virtual camera at a distance of $d_{\rm camera}=10^4 r_{\rm g}$ and appropriately re-scale the flux density $\sim1/r^2$ to the distance of Sgr A* -- here assumed to be 8.127 kpc \citep{Reid2019}. The field of view of the camera is set to $44 r_{\rm g}$, and we sample the image with $400^2$ pixels. 

\subsection{Thermodynamics}

Since the GRMHD simulations only provide information on the dynamically dominant ions, we must introduce parametrizations for the electron properties. First, assuming a pure hydrogen gas, charge-neutrality of the plasma dictates that $n_{\rm p}=n_{\rm e}$, where $n_{\rm p}$ is the proton number density and $n_{\rm e}$ is the electron number density. Secondly, we compute the electron temperature as

\begin{align}
    \Theta_e= \frac{ m_{\rm p} P}{m_{\rm e} \rho (1+R)},
\end{align}

$R$ then sets the ratio between the electron to proton temperature \footnote{here we adopt a adiabatic index of $\hat{\gamma} = 4/3$. Therefore, our formula deviates from Eq. (7) in \cite{CollaborationAkiyamaEtAl2019d}}. For this ratio, we follow the prescription of \cite{MoscibrodzkaFalcke2016a} by setting
\begin{align}
    R= R_{\rm high}\frac{\beta^2}{1+\beta^2} + \frac{1}{1+\beta^2}\label{eq:rhigh},
\end{align}
where $\beta=P_{\rm gas}/P_{\rm mag}$ is the ratio of thermal pressure $P_{\rm gas}$ and magnetic pressure $P_{\rm mag}$. Via this prescription, we can vary the uncertain emission location using the factor $R_{\rm high}$, where low values of $R_{\rm high}$ lead to relatively hot disk electrons promoting the emission from the disk region. Conversely, large $R_{\rm high}$ values lead to emission from the more magnetized regions near the jet wall. The following values are adopted in our study: $R_{\rm high}\in\{1,10,40,160\}$. As the simulations are all scaled to the same average 230 GHz flux density by changing the density normalization of the GRMHD data, increasing $R_{\rm high}$ leads to larger plasma densities. This is because larger values of $R_{\rm high}$ correspond to colder disk electrons which lower mm wavelength emission, consequently a larger number is required to arrive at a given 230 GHz flux density.  It is important to keep density normalization in mind when interpreting the results as it has consequences e.g., on the Faraday depth and polarization degree of the solutions.   

During a magnetic flux eruption, particle acceleration due to magnetic reconnection can alter the shape of the electron distribution function (eDF) \citep{RipperdaLiskaEtAl2022,Hakobyan2023,ZhdankinRipperdaEtAl2023}. Non-thermal particles gain importance in radio- and IR frequencies \citep{Ozel2000,Davelaar2018} and are necessary to explain the inverted IR spectrum during strong flares \citep{DoWitzelEtAl2019} as well as the highest X-ray fluxes \citep{HaggardNynkaEtAl2019}. However, how and where to assign non-thermal eDFs in our GRMHD domain is very uncertain; parametrizations of the eDF can be done in various ways, which would enlarge our parameter space substantially. In this work, as a first step, we will hence only discuss mm-emission which is caused by the thermal core of the eDF and postpone a full eDF study to future work. 

Lastly, the GRMHD simulations which are run in dimensionless geometric units ($G=M=c=1$) under a ``test-fluid'' assumption (neglecting the contribution of the plasma to the gravitational field) need to be scaled to concrete physical units. Thus we introduce scaling factors for plasma mass, length, and time. Units of length are scaled with ${\mathcal L}=r_{\rm g}$. Units of time are scaled with ${\mathcal T} = r_{\rm g}/c$, and the mass unit is set by the scaling factor ${\mathcal M}$. The mass unit scales plasma density via ${\mathcal M}/{\mathcal L^3}$ and is related to the mass accretion rate via $\dot{M} = \dot{M}_{\rm sim}{\mathcal M}/{\mathcal T}$, where $\dot{M}_{\rm sim}$ is the mass accretion rate in geometric units. 
The black hole mass is tightly constrained by observations of stellar orbits in the galactic centre; see, e.g., \cite{Do2019,GRAVITY2019}. Consequently, we set the black hole mass to $M=4.14\times10^6 M_\odot$ which fixes $\mathcal{L}$ and $\mathcal{T}$.
Since the plasma density is poorly constrained, for each parameter combination $(R_{\rm high},i,a_{*})$, we use ${\mathcal M}$ as a free parameter to match the average flux density of our models to the observed flux density of $2.4~\rm Jy$ \citep{EventHorizonTelescopeCollaborationAkiyamaEtAl2022}.

\section{Results}\label{sec:results}

\subsection{Polarization loops during flux eruptions}\label{sec:mad}
To study the effect of flux eruptions on the polarization signatures, we first analyze the counter-rotating $a_*=-0.9375$ MAD simulation since it features the strongest and most frequent flux eruptions in our sample. The radiation transfer parameters are set to $i=10^{\circ}$ and $R_{\rm high}=1$ such that we look nearly down the black hole spin axis and the emission is disk-dominated. We monitor the evolution of the polarized flux in the $\mathcal{Q-U}$ plane over a sliding time window of $180 ~GM/c^2\simeq 1$ hour, similar to the period observed by \cite{wielgus22}.   
\begin{figure*}
    \centering
    \includegraphics[width=0.8\textwidth]{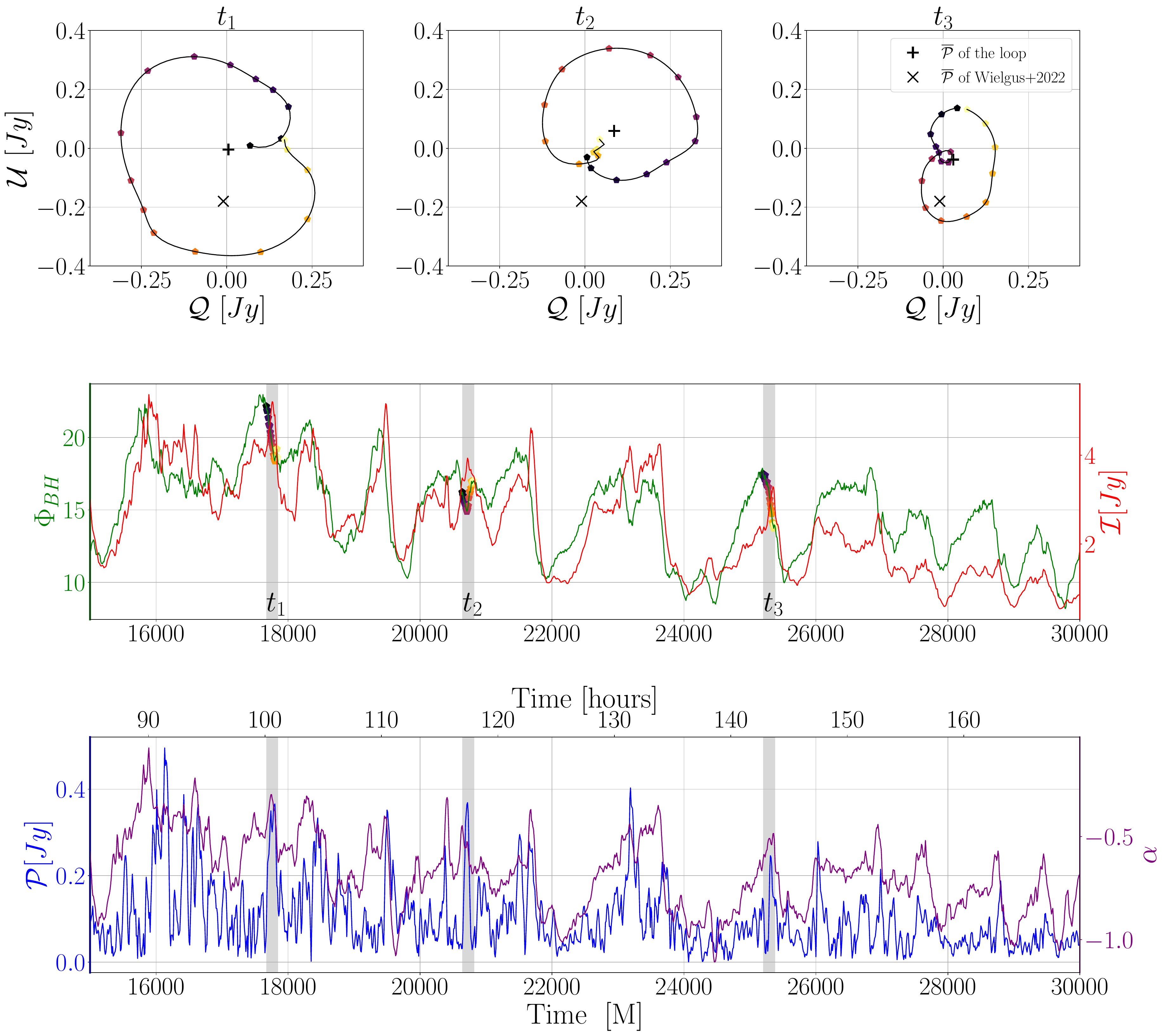}
    \caption{\textit{First row}: emerging loops in $\mathcal{Q}-\mathcal{U}$ diagram at three different times. The colours represent the time, such that darker colours show the earlier times and brighter colours show the later times; hence the loops rotate counter-clockwise. The interval between consecutive data points is $200$ seconds. We also mark the mean polarization as '+'  and indicate the observed one as 'x' \citep{wielgus22}. \textit{Second row}: Comparison of magnetic flux $\Phi_{\rm BH}$ with the total intensity $\mathcal{I}$ while marking the loop instances at each time interval. \textit{Third row}: linear polarization where $|\mathcal{P}| = \sqrt{\mathcal{Q}^2 + \mathcal{U}^2}$.and $213.1 - 229.1$ GHz spectral index $\alpha$.  Specific simulation parameters are $a_{*} = -0.9375$, $R_{\rm high}=1$, $i=10^{\circ}$, MAD.  See \url{https://doi.org/10.5281/zenodo.8302230} for an animation of this simulation.}
    \label{fig:loops-lightcurves-mad} 
\end{figure*}
The cadence is set by the output frequency of the GRMHD simulations to $10 ~GM/c^2$, which results in sufficient data points to see coherent structures in the polarization time domain. 
Three exemplary loops associated with flux eruptions are shown in the top panel of Figure~\ref{fig:loops-lightcurves-mad}. All selected cases show periods of $\simeq 1$  hour and move in a counter-clockwise direction. We recover a range of polarized flux from $0.2-0.3$ Jy resulting in varying loop radii. While the ALMA loop was significantly off-centre at $(-0.19,0.01)~\rm Jy$ (as indicated by x on the figure), all our loops are centred close to the origin of the $\mathcal{Q-U}$ plane. Our $t_3$ loop, similar to the ALMA loop, recovers a secondary 'tiny loop' close to the origin; see \cite{VosMoscibrodzkaEtAl2022} on an explanation of how such features arise in hotspot models. Apart from an overall flip (which can be obtained by instead observing at $i=170^\circ$), and offset, this ``Prezel''-shaped loop qualitatively shows similar behaviour to the ALMA loop. 

We find a correlation between the formation of the loops and the slope of the magnetic flux; see for full details Appendix A. As magnetic flux is being accumulated at the event horizon (a positive slope, $\frac{d}{dt} \phi_{BH}>0$), prominent large polarization loops are rare. However, flux eruptions (a negative slope, $\frac{d}{dt}\phi_{BH}<0$) are almost always accompanied by loops in the $\mathcal{Q-U}$ plane. 
It is worth noting that the ALMA loop occurred on the rising slope of flux density (corresponding to the re-accretion phase) where loops are less pronounced in our simulations.  

In the second panel of Figure~\ref{fig:loops-lightcurves-mad}, we simultaneously monitor the evolution of $230$ GHz flux with the horizon penetrating magnetic flux $\Phi_{\rm BH}$. 
Overall, the data shows $13-15$ large flux eruptions characterized by a sudden drop in $\Phi_{\rm BH}$ which results in the typical recurrence time of $\approx 1000 \rm M$.  
Generally, the post-eruption radiation flux is lower than the flux in the build-up phase. This behaviour was also noted by \cite{JiaRipperdaEtAl2023}.  However, we also find several instances where the $230$ GHz flux density sharply peaks during flux eruption.  In our sample, the ``peaking'' flux eruptions occur especially in earlier times where the intensity and optical depth are also the highest.  

\begin{figure*}
    \centering
    \includegraphics[width=\textwidth]{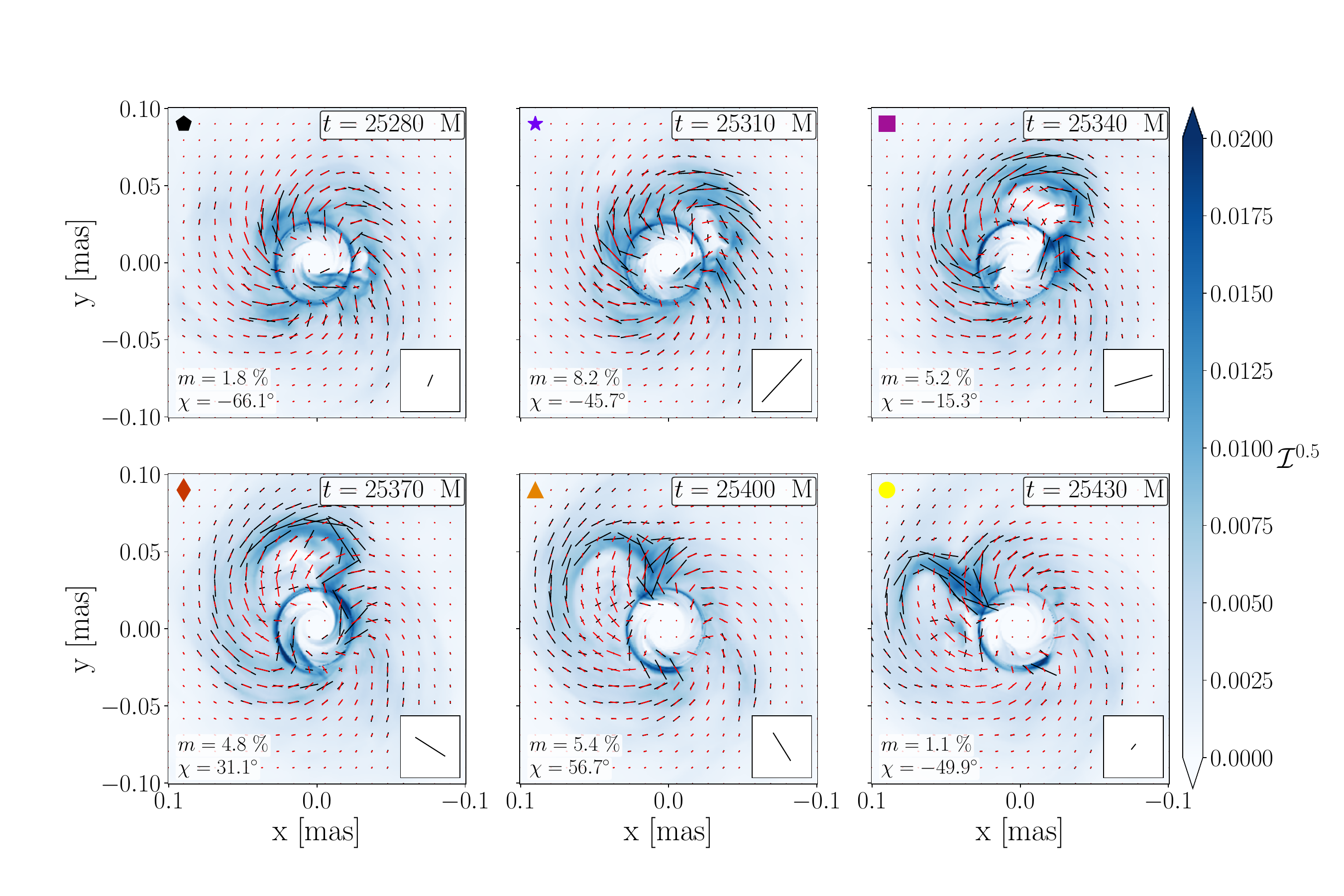}
    \caption{Intensity map for the loop at $t_3$ capturing the flaring event for the fiducial case (MAD, $a_* = -0.9375$, $R_{\rm high} =1$, $i=10^{\circ}$). The overlayed ticks show the fraction of the polarization flux magnitude, positioned by the polarisation angle; the black tick represents the instantaneous electric vector while the red ticks are the average electric vector. The boxes on the bottom right of each panel represent the average electric vector angle of that panel ($\chi$). In addition, we have included the degree of polarisation of each panel $m = \frac{\overline{\mathcal{P}}}{\overline{\mathcal{I}}}$. Colored symbols facilitate comparison with Figure \ref{fig:QU-timeseries} which shows the time-series of the fluxes.  }
    \label{fig:grrt_multipanel}
\end{figure*}

The intensity variations can be attributed to the evolution of the magnetic flux eruption, as illustrated in Figure~\ref{fig:grrt_multipanel}: during the expulsion of magnetic flux (hence for $\frac{d}{dt}\Phi_{BH}<0$), a cavity of low density plasma appears in the image at the photon ring (top-left panel).  As more flux escapes from the black hole, the cavity grows in size and starts interacting with the accretion flow (top row, second and third panel). Disk material is compressed against the flux bundle's surface leading to a locally enhanced emission (yielding the distinct peaks in intensity seen in panel 2 of Figure~\ref{fig:loops-lightcurves-mad}). Since the flux bundle rotates slower than the disk, the compression occurs in the back of the flux bundle.  Compression of the plasma also leads to a change in the EVPA compared to the time average values (cf. black vs. red ticks).  
As the flux bundle expands in the ambient accretion flow, the region of enhanced emissivity adiabatically cools and becomes sub-dominant (bottom row). Disk material is pushed away, leaving behind a large cavity of under-luminous (but hot) jet plasma (bottom right panel).  In our simulation, this emission 'hole' is responsible for the lower post-flare radiation flux. 
The EVPA has now performed a full loop while the flux bundle has progressed by roughly half of a circle.  
As the hole is dissolved, the intensity recovers to pre-flare levels on a range of timescales ($1-5 ~\rm hrs$) accompanied by re-accretion of the magnetic flux.  
It is interesting to note that a similar recovery of the mm-flux on a timescale of $\sim 2~\rm hrs$ was also observed after the X-ray flare on 11. April 2017 \citep{WielgusMarchiliEtAl2022}.  

We further monitor the linearly polarized flux $\mathcal{P}$ and spectral index $\alpha$ where $F_{\nu} \propto \nu^{\alpha}$ in panel 3 of Figure~\ref{fig:loops-lightcurves-mad}. 
Both quantities, but $\alpha$ in particular, correlate with the magnetic flux.  The evolution of spectral index and polarization degree indicates a changing optical depth, consistent with a transition from optically thick to optically thin emission during the flux eruption event.  In the aftermath, as the inner accretion disk is filled back with emitting plasma, intensity and spectral index climb back to pre-flare values.  
Similar trends, that is a spectral index decrease by $0.5\pm0.2$ across the mm peak as well as an excess of linear polarization during the flare (increasing from 9\% to 17\%) were also observed by \cite{MarroneBaganoffEtAl2008} in their lightcurve of the 2006 simultaneous X-ray and mm wavelength flare.  Since the cooling time for electrons emitting in the mm band is much longer than the mm-flare duration, this evolution must be driven by adiabatic expansion of the flaring region, just like in our simulations which neglect radiative cooling.  

Regarding the ALMA loop with data taken just $\simeq 1~\rm hr$ after the X-ray peak, it is noteworthy that no sub-mm flare was recorded on April 11, rather the $230~\rm GHz$ lightcurve starts at a low value of $\simeq 1.8\rm Jy$ followed by a recovery to usual quiescent values $\sim2.5~\rm Jy$.  At the same time, the spectral index increases from $\simeq -0.2$ back to its quiescent value which is consistent with zero.  In the context of our simulations, one interpretation of this spectral timing behavior is that the ALMA loop was observed during the re-accretion phase characterized by increasing flux and spectral index on a timescale of hours (cf. the large flux eruption event in Figure \ref{fig:loops-lightcurves-mad} before $20\, 000~\rm M$).  
An alternative interpretation to the recovery of the mm-flux density during the ALMA loop is that intense heating during the X-ray flare (coincident with the flux eruption) has shifted the characteristic frequency of the emitting electrons out of the mm-band.  The observed recovery is then governed by the radiative cooling of the X-ray and IR emitting electrons.  If the cooling time is fast enough this could lead to an increase of mm-flux during the ``loopy phase'' just after the flux eruption.  This scenario should be tested with future simulations that take into account radiative cooling.  
\\

\begin{figure}
    \centering
    \includegraphics[width=0.5\textwidth]{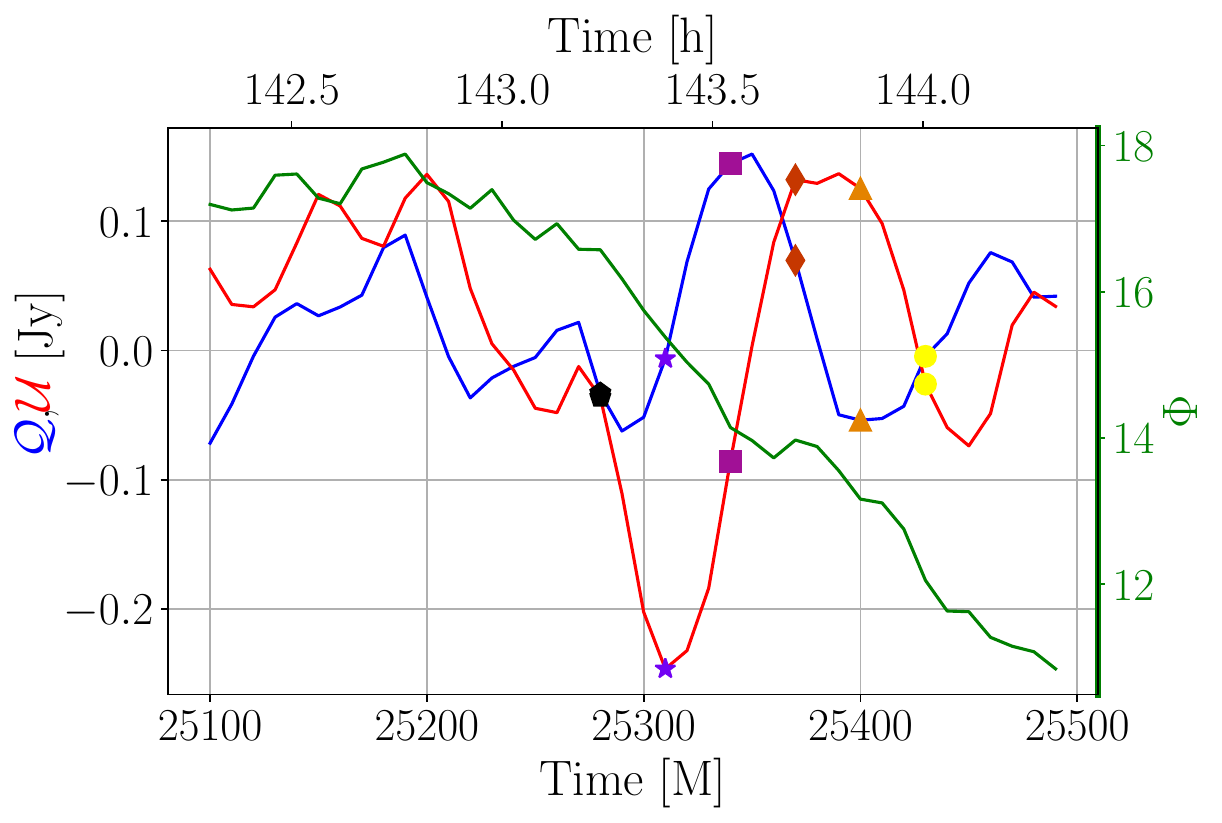}
    \caption{Comparison of Stokes $\mathcal{Q}$, Stokes $\mathcal{U}$, and magnetic flux {$\Phi_{BH}$} for a loop formed between $25200 M$ and $25380 M$ ($t_3$ in Figure~\ref{fig:loops-lightcurves-mad}), for the case of a black hole with spin $a_* =-0.9375$, MAD disk with $R_\text{high} =1$ and inclination angle $i=10^{\circ}$. The markers corresponds to each panel of Figure \ref{fig:grrt_multipanel}}
    \label{fig:QU-timeseries}
\end{figure}

The behaviour of the $\mathcal{Q-U}$ parameters during flaring events can also be seen in the time series of these fluxes. Focusing on the loop formed at $t_3$, in Figure~\ref{fig:QU-timeseries}, we show the time evolution of the linear polarisation components $\mathcal{Q}$ and $\mathcal{U}$ in comparison with the magnetic flux. A telltale sinusoidal pattern emerges immediately after the local peak of the magnetic flux, which corresponds to the loopy feature in the $\mathcal{Q-U}$ plane.  Identification of such features in polarized lightcurves should be fairly straightforward.  

\subsection{Inclination dependence}\label{sec:inclination}

\begin{figure}
    \centering
    \includegraphics[width=0.4\textwidth]{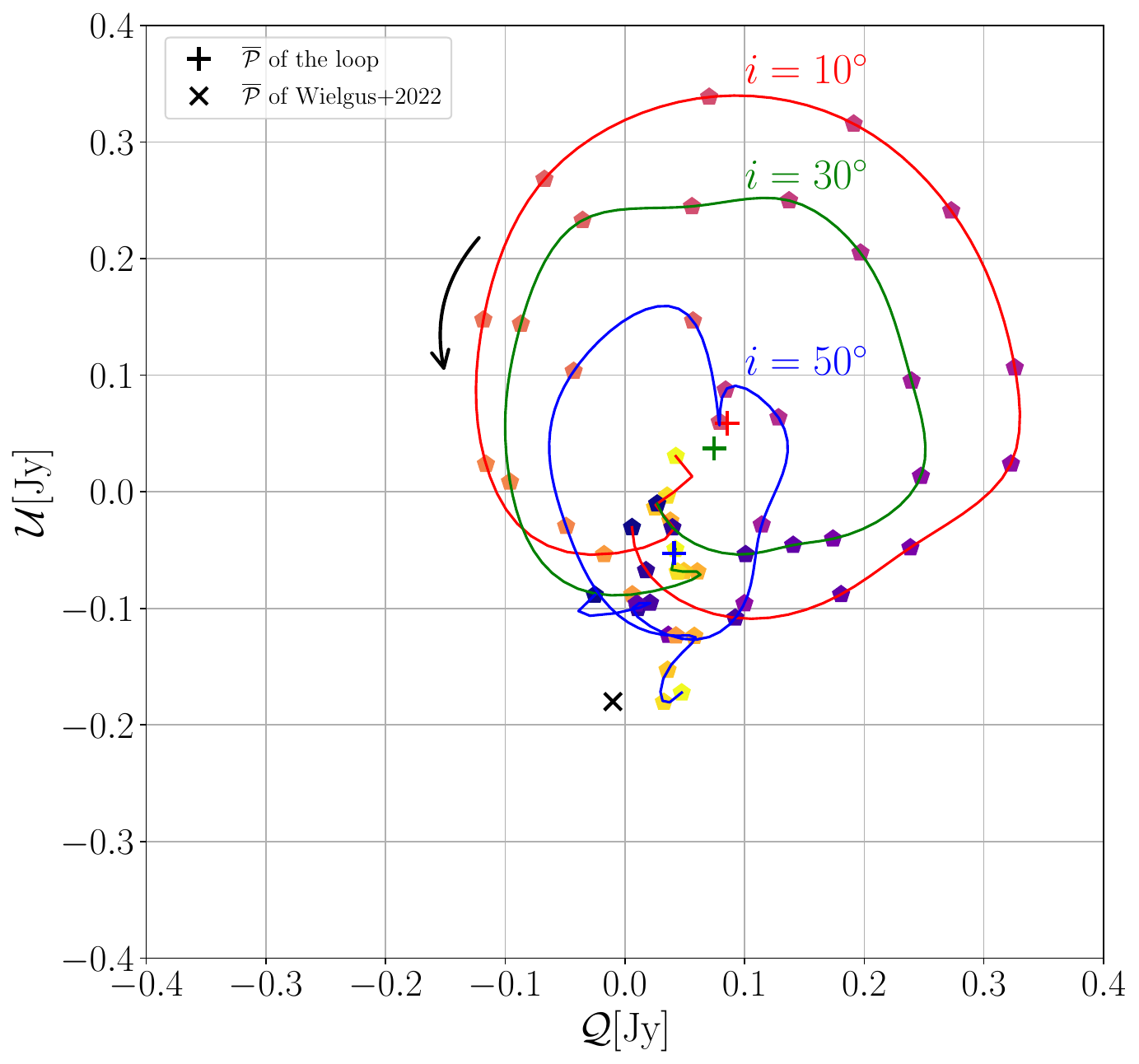}
    \caption{Inclination dependence of the loopy feature formed in $t_2$ as in Figure \ref{fig:loops-lightcurves-mad}. (MAD, $a_*= -0.9375$, $R_\text{high}=1$)}
    \label{fig:loop-incl}
\end{figure}

To test whether the observation of the rotating EVPA can place constraints on the inclination, we investigate the dependence of the polarization loops on the viewing angle. To this end, in Figure~\ref{fig:loop-incl}, we show the loop formed in $t_2$ for three different inclinations with respect to the black hole spin axis: $10^\circ,30^\circ$, and $50^\circ$. Due to the loss of symmetry, the time-averaged polarization degree typically increases with larger viewing angles, however, in our simulation this does not lead to an additional offset of the polarization loops from the origin.  In fact, the loop-averaged polarization is even slightly closer to the origin for larger inclinations.  
This can be interpreted in the following way: during the flux eruption, the polarized emission from the loop boundary (see also Figure~\ref{fig:grrt_multipanel}) is dominant over any background ``shadow'' contribution which does not significantly offset the average polarization.  
The size of the loops shrinks as we incline the line of sight; for inclinations larger than 50 degrees, the linear polarisation components show more stochastic behaviour, and loops are less clear.  
The loss of coherent loops for larger viewing angles can also be quantified by the corresponding decrease of correlation between $\qstokes$ and $\ustokes$ discussed further in Section \ref{sec:timing}.  

\subsection{Polarization loops in non-flaring SANE simulations}\label{sec:sane}

\begin{figure*}
    \centering
    \includegraphics[width=0.8\textwidth]{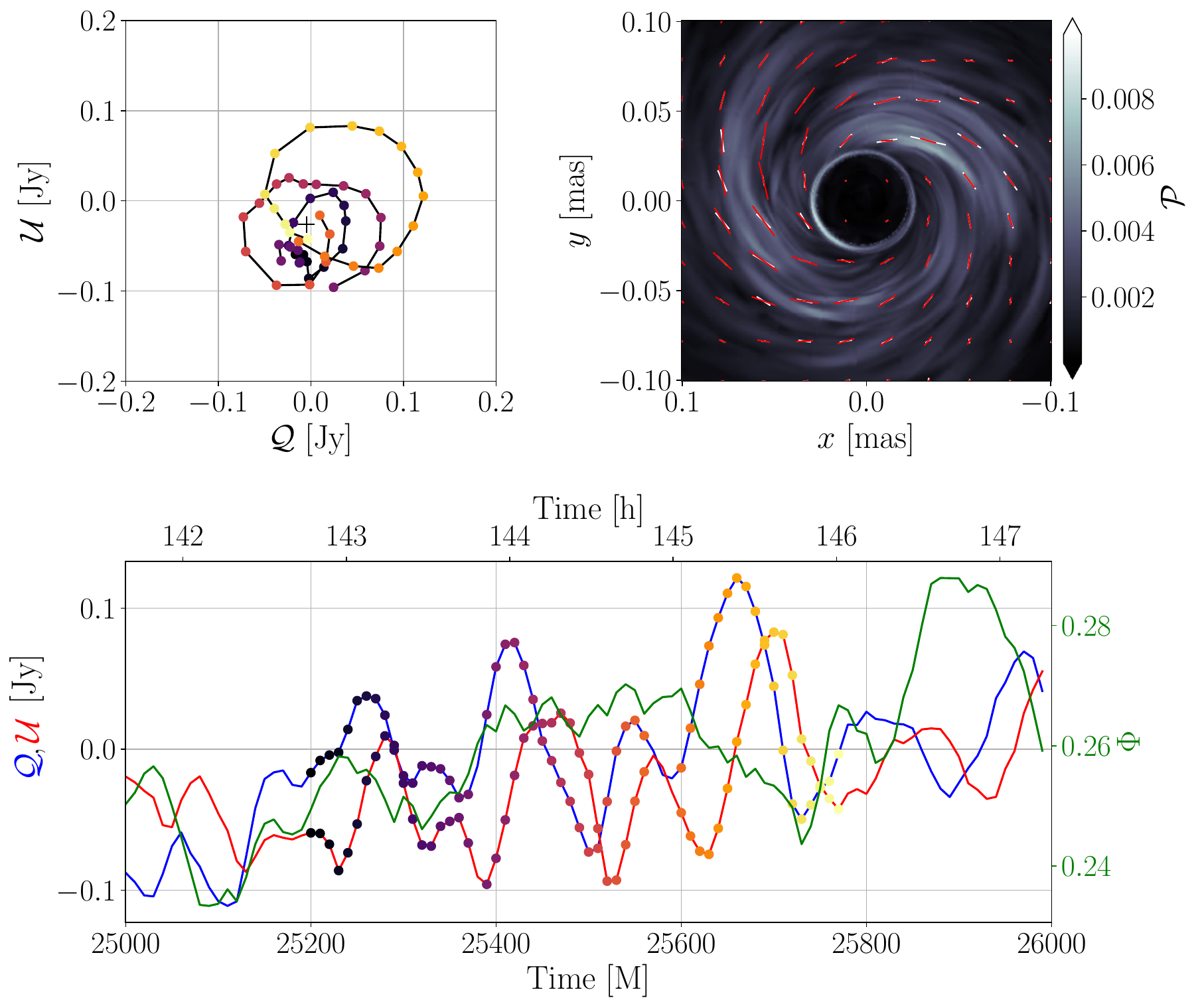}
    \caption{Consecutive loops in the time interval of 25200-25780 M. The second panel displays an instantaneous snapshot of linearly polarised flux at 25400 M, where the red ticks represent the average electric vector and the white ticks are the instantaneous electric vector. Second row: the time series of $\mathcal{Q}$ and $\mathcal{U}$ corresponding to the loops in the first panel. The colour of the dots represents the time. (SANE, $a_*=-0.9375$, $i=10^\circ$, $R_{\text{high}}=1$, See \url{https://doi.org/10.5281/zenodo.8302230} for an animation of this simulation)}
    \label{fig:multi_sane_xl} 
\end{figure*}

In the previous section, we show that the loops generally form during magnetic flux eruptions; this implies that the loops in the $\mathcal{Q}$ and $\mathcal{U}$ plane could provide evidence of Sgr A* being in a MAD state. To test this hypothesis, we analyse our SANE simulations, which do not exhibit magnetic flux eruptions, as the magnetic flux saturates at a much lower value of around $\Phi_{\rm BH}\sim 0.25$. In contrast to our hypothesis, the SANE $R_{\rm high}=1$ model does form loops. This behaviour can be seen in Figure~\ref{fig:multi_sane_xl}; on the top-left panel, several consecutive loops with a period of $\sim 1$ hour are visible. Another aspect of this behaviour can be traced in the bottom of Figure~\ref{fig:multi_sane_xl} where the time series of the $\mathcal{Q,U}$ (blue and red lines) is overplotted with the magnetic flux (green line). Multiple sinusoidal patterns are visible in $\mathcal{Q,U}$; however, the magnetic flux does not exhibit substantial dissipation events, in contrast to the MAD models. This indicates that the loops must have a different physical origin. 

To elucidate the origin of these features, in the top right panel of Figure~\ref{fig:multi_sane_xl} we show the instantaneous map of the polarized flux ${\mathcal{P}}$ at 25400 M, overlayed with the instantaneous electric vector (white) and time-average electric vector (red). In this snapshot, there is only a very small deviation between the average and instantaneous angle of the electric vector. However, the lengths of the white ticks change since the linear polarisation is not constant as a function of time. The polarized intensity shows an enhancement of emission due to azimuthal $m=1$ and $m=2$ modes (such that emission is enhanced going once respectively twice around the origin), resulting in an asymmetry in the emission morphology. As these  $m=1$ and $m=2$ modes orbit around the black hole they change the orientation of the locally emitted EVPA in a periodic fashion, similar to the orbiting flux tube in the MAD case. However, in contrast to the flux tubes in the MAD case, the spirals are longer-lived, therefore generating multiple consecutive loops in the $\mathcal{Q-U}$ plane. Additionally, due to the lower contrast of the emitting features, the resulting net polarization is smaller than in the MAD case. 

When moving the emission away from the disk when $R_{\rm high} \geq 10$, the net polarization decreases further. Additionally, the trajectories in the $\mathcal{Q-U}$ plane becomes significantly slower and stochastic: in this case, we do not obtain any polarization loops consistent with the observations, see Appendix \ref{sec:Rhigh} for more details. This behaviour is not seen in MAD simulations which show loops for any of the tested $R_{\rm high}$ parameters since the emission in MADs is generally less sensitive to $R_{\rm high}$.

\subsection{Timing properties of {Q-U} loops}\label{sec:timing}

\begin{figure*}
     \centering
     \begin{subfigure}[b]{0.33\textwidth}
         \centering
         \includegraphics[width=\textwidth]{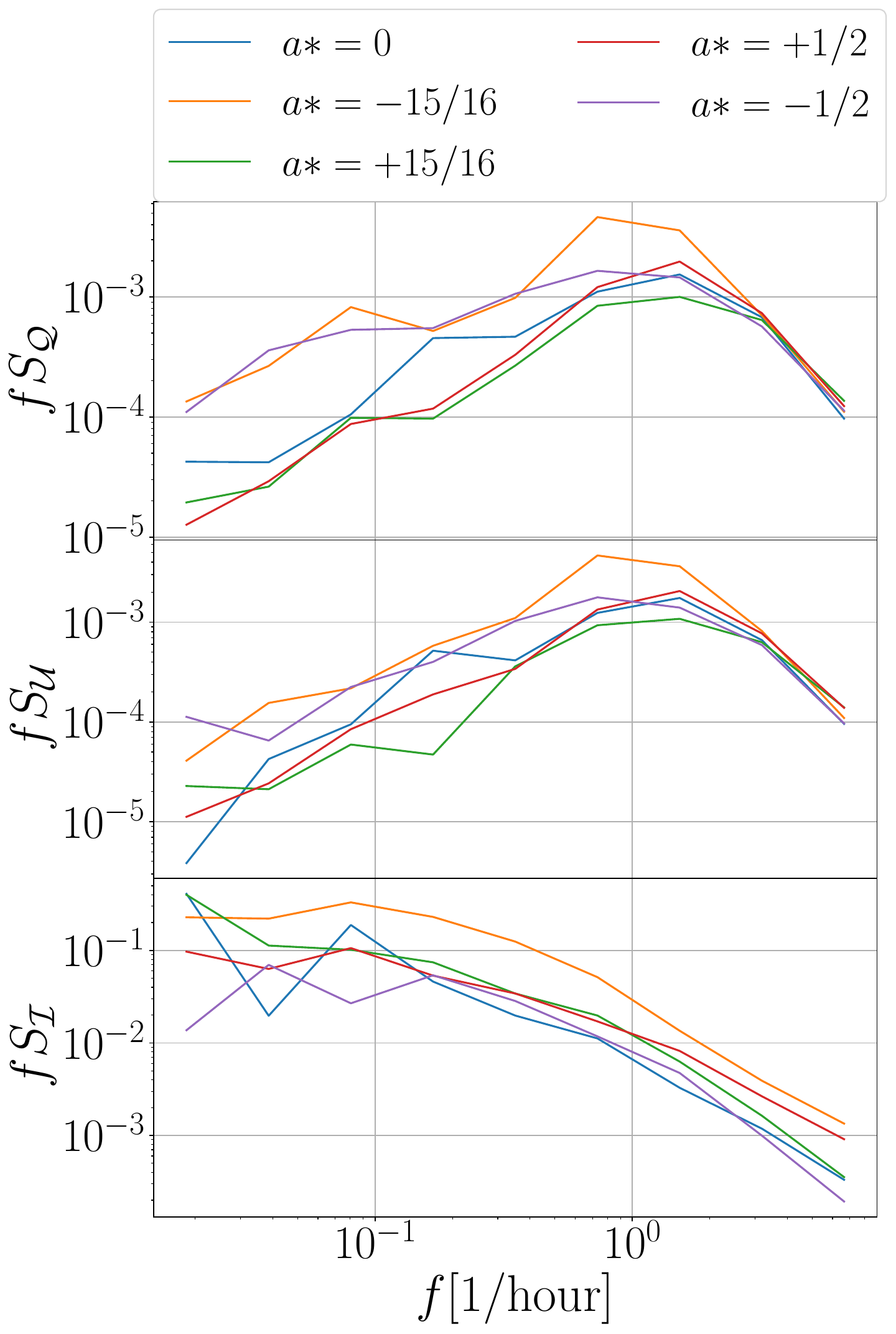}
         \caption{MAD, $R_{\text{high}}=1$}
         \label{fig:psd-MAD}
     \end{subfigure}
     \hfill
     \begin{subfigure}[b]{0.33\textwidth}
         \centering
         \includegraphics[width=\textwidth]{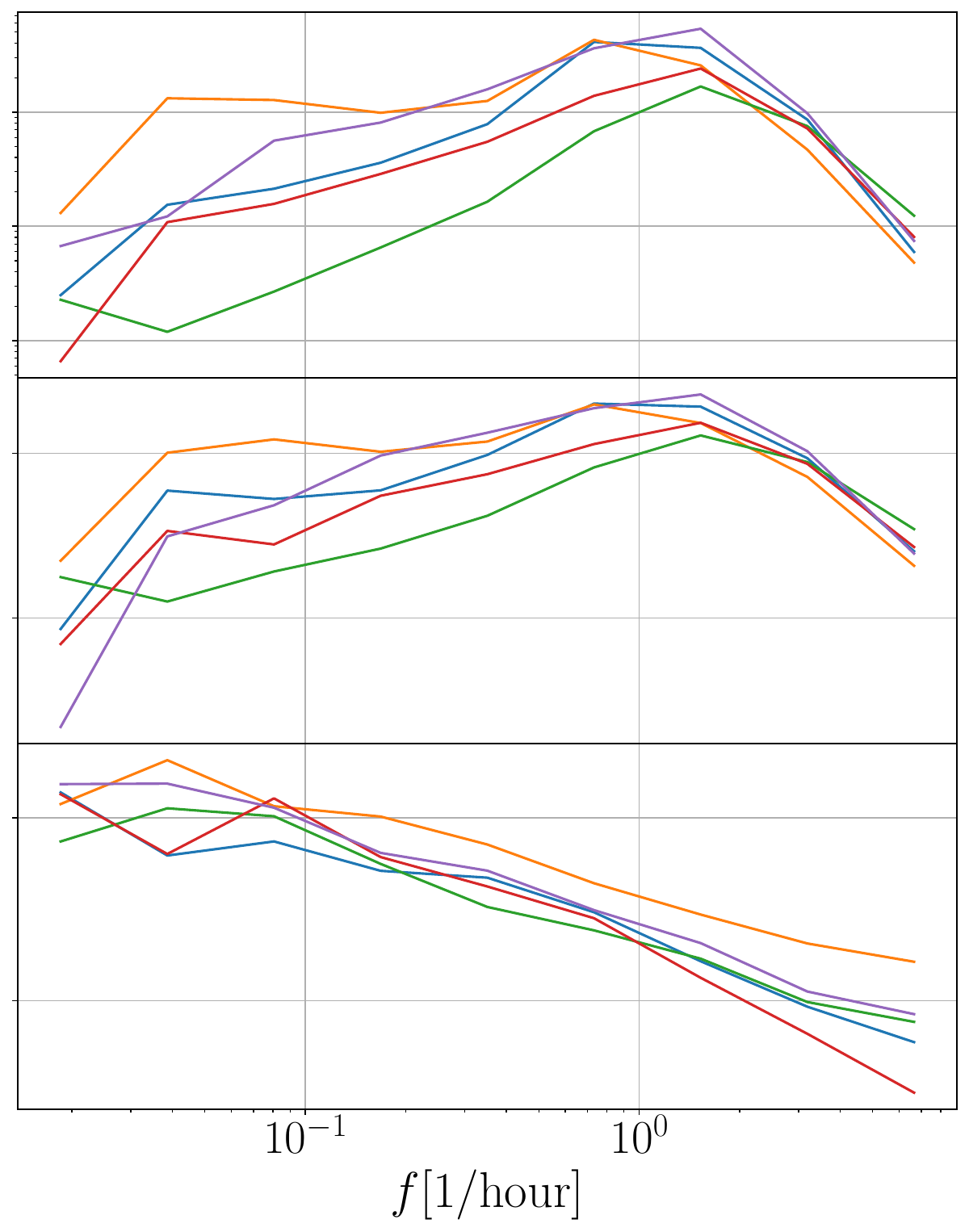}
         \caption{SANE, $R_{\text{high}}$ = 1}
         \label{fig:psd-SANE}
     \end{subfigure}
     \hfill
     \begin{subfigure}[b]{0.33\textwidth}
         \centering
         \includegraphics[width=\textwidth]{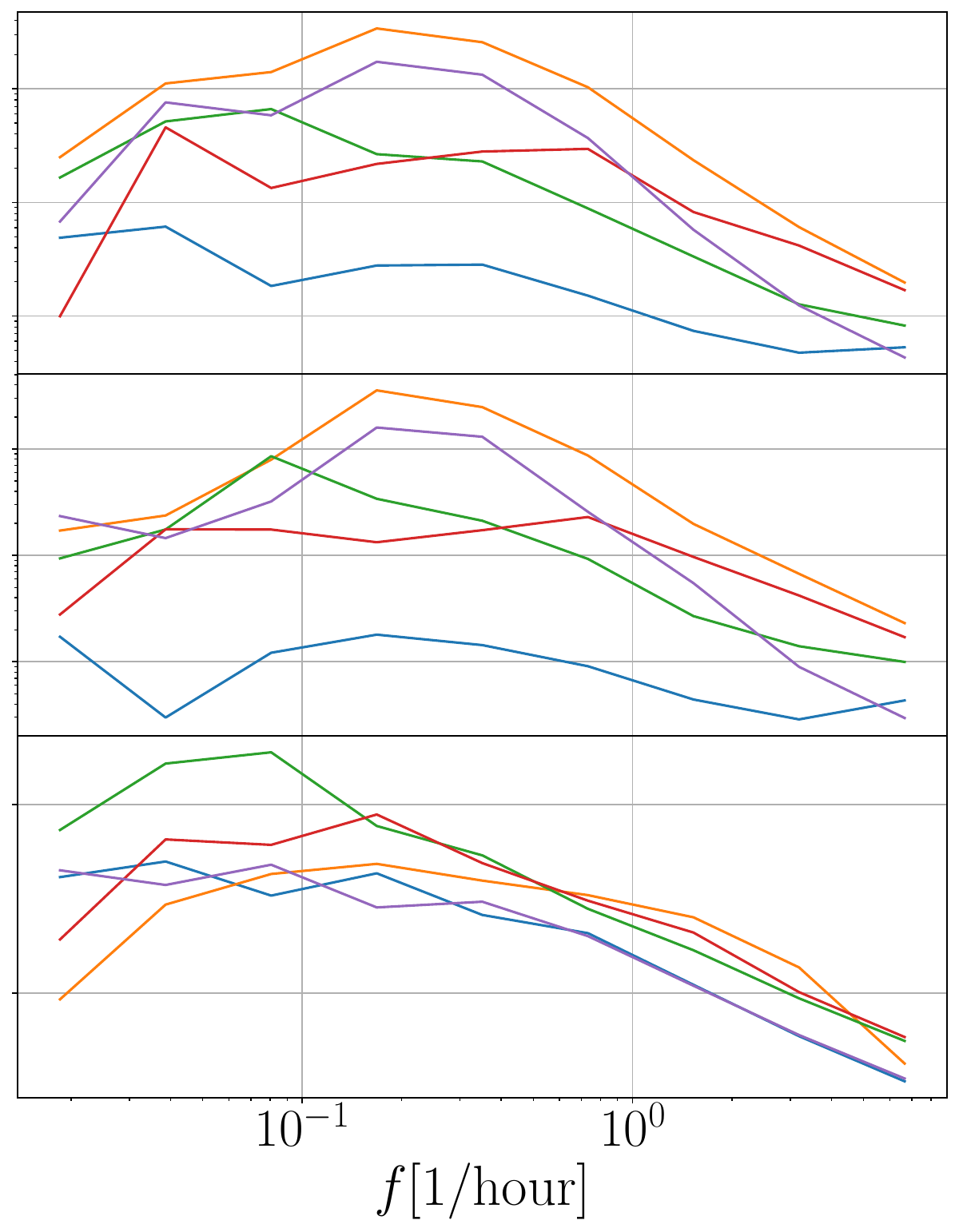}
         \caption{SANE, $R_{\text{high}}=160$}
         \label{fig:psd-SANE-160}
     \end{subfigure}
        \caption{The power spectrum densities of three different disk configurations with inclination $i = 10^{\circ}$. For the case of hot disk $R_{\text{high}}=1$, the stokes parameter $\mathcal{Q}$ and $\mathcal{U}$ show a peak at $f = 1 $[1/h] while the $\mathcal{I}$ follows the red noise variability. For the case of the cold disk ($R_{\text{high}}=160)$, the emission is coming from the jet which has less variability. Consequentially, no variability excess is obtained in the case of jet dominated emission with non-rotating black hole ($a_*=0$).  }
        \label{fig:three-psds}
\end{figure*}

To further analyze the systematic signal in the $\mathcal{Q}$ and $\mathcal{U}$ time series, we compute power spectrum densities of $\mathcal{I}, \mathcal{Q}$ and $\mathcal{U}$. These are shown in Figure~\ref{fig:three-psds} for both MAD and SANE disks with different spins at $i=10^{\circ}$.  
For the case of a hot disk ($R_{\rm high} = 1$), at around $f = 1 \ \rm [1/h]$ a broad peak emerges for $\mathcal{Q}$ and $\mathcal{U}$ PSDs while the power spectrum for $\mathcal{I}$ follows the typical red-noise accretion variability spectrum \cite[e.g.][]{HoggReynolds2016,PorthChatterjeeEtAl2019a,BollimpalliMahmoudEtAl2020}. 

This qualitative behaviour is generally present for both MAD and SANE disks, however in the cold-disk SANE case with $R_{\rm high} = 160$ the $\mathcal{Q},\, \mathcal{U}$ PSDs are broader and peak at much lower frequencies $\sim 0.3/\rm h $. This can be explained by the emission site: for SANE, $R_{\rm high} = 160$ the emission is located at larger distances in the jet, corresponding to slower variations with less systematic rotation compared to the disk mid-plane.  Accordingly, for the zero spin case with $R_{\rm high} = 160$, we do not observe any significant periodicity due to the absence of jet rotation.  By contrast, since $R_{\rm high}$ has less impact on the emission site in MAD disks which are highly magnetized also in the mid-plane \cite[c.f. Figure 4 of][]{CollaborationAkiyamaEtAl2019d}, the PSDs for MAD disks are insensitive to the values of $R_{\rm high}$ explored in our work.  

\begin{figure}
     \centering
     \begin{subfigure}[b]{0.5\textwidth}
         \centering
         \includegraphics[width=\textwidth]{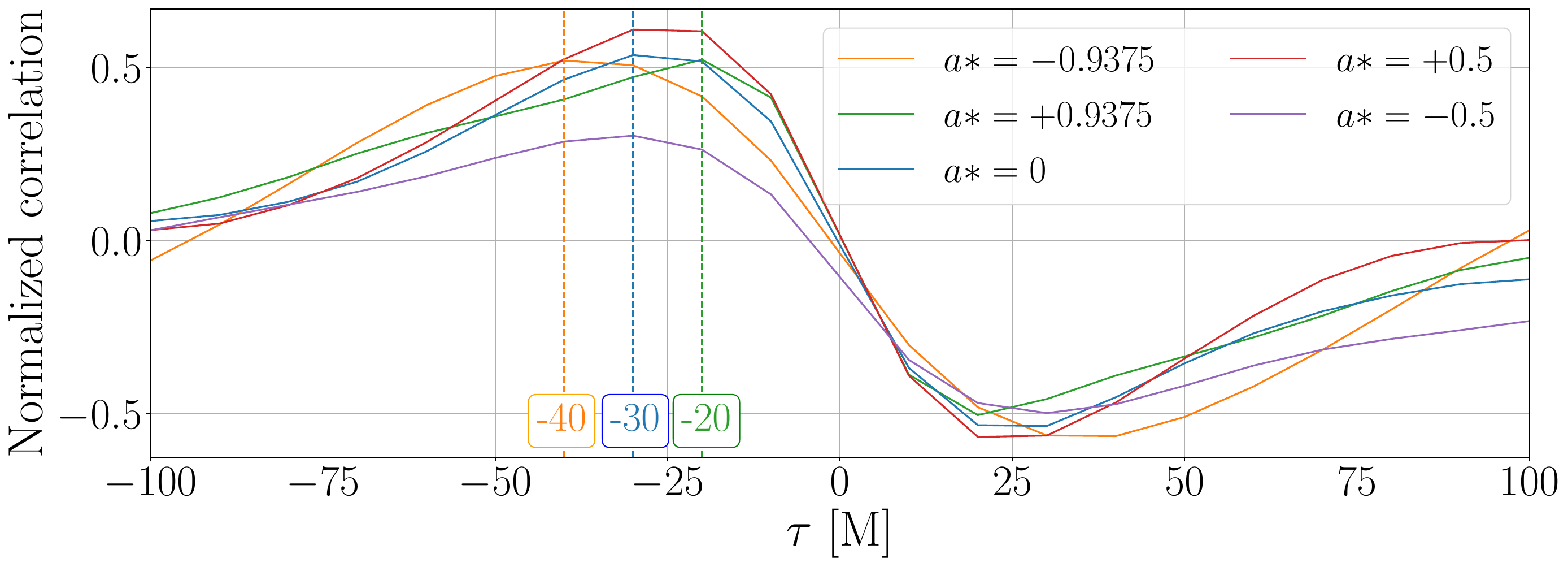}
         \caption{MAD}
         \label{fig:ccMAD}
     \end{subfigure}
     \hfill
     \begin{subfigure}[b]{0.5\textwidth}
         \centering
         \includegraphics[width=\textwidth]{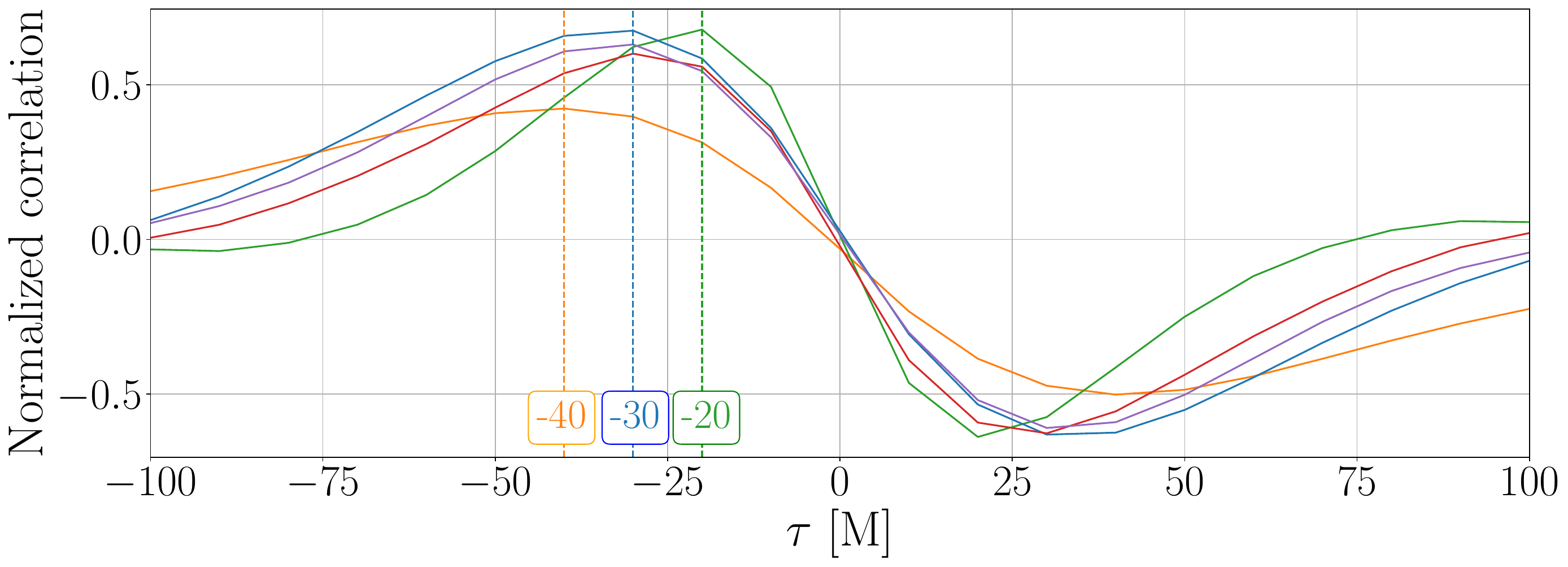}
         \caption{SANE}
         \label{fig:ccSANE}
     \end{subfigure}
        \caption{The cross-correlation between $\mathcal{Q}$ and $\mathcal{U}$. The time lag corresponding to the maximum correlation is marked for each case. The colours represent different spins. For all cases $R_{\rm high}=1$ and $i =10^{\circ}$. }
        \label{fig:ccQU}
\end{figure}

To further quantify the structural variability, we compute the normalized cross-correlation between $\mathcal{Q}$ and $\mathcal{U}$ for models with various spins.  We use the {\tt Scipy} Signal package to correlate the quantities 
\begin{align}
    \mathcal{Q}_{\rm normal} = \frac{\mathcal{Q}-\overline{\mathcal{Q}}}{N\sigma_{\mathcal{Q}}}\, ; \hspace{1cm}
    \mathcal{U}_{\rm normal} = \frac{\mathcal{U}-\overline{\mathcal{U}}}{\sigma_{\mathcal{U}}}\, 
    \label{eq:qnormal}
\end{align}
where $\overline{\mathcal{Q}}$ and $\overline{\mathcal{U}}$ are the time averages, $N$ is the number of samples and $\sigma_{\mathcal{Q}}$, $\sigma_{\mathcal{U}}$ are the standard deviations of the fluxes.  
Figure~\ref{fig:ccMAD} and \ref{fig:ccSANE} depict the cross-correlations of MAD and SANE disks. For both disk configurations, the cross-correlation is anti-symmetric about the origin, and the maximum and minimum occur at similar time lags for each spin -- regardless of the disk being MAD or SANE. 
The sense of rotation is indicated by the slope of the cross-correlation through the origin, all our simulations feature counter-clockwise rotation and are thus sensitive to the rotation of the disk, not the black hole.  
For the co-rotating case $a_*=+0.9375$ the maximum cross-correlation is at $\tau = -20~\rm M$ and for the negative spin $a_* = -0.9375$ it is at $\tau = -40 ~\rm M$. This implies that the linear polarisation flux components vary with a time delay of 7 respectively 13 minutes.  
Interpretation of this result is straightforward if we model $Q\sim \sin(2\pi t/P)$ and $U\sim\cos(2\pi t/P)$ (cf. Figure~\ref{fig:QU-timeseries}): at zero time-shift $\tau=0$, we obtain a loop with minimal cross-correlation; once $U$ is shifted by a quarter period $\pm P/4$, both signals vary in-phase/anti-phase. While the peaks are fairly broad, this implies a \textit{typical} periodicity of 27 to 54 minutes.  
The universal behavior of the cross correlations shows that \qstokes-\ustokes~ loops are a robust prediction for low inclination MAD and $R_\text{high}=1$ SANE simulations.  
Upon increasing the inclination, the correlation of $\mathcal{Q}$ and $\mathcal{U}$ decreases: for example, while the maximum normalized cross-correlation for the MAD $a^*=0.5$ run with $i=10^\circ$ is $\simeq 0.6$, it decreases to $\simeq 0.5$ for $i=30^\circ$ and to $\simeq 0.2$ for $i=50^\circ$.  
Our analysis shows that the SANE $R_{\rm high}=1$ case shows a similar dependence on inclination. This is consistent with the loss of polarization loops for increasing inclination discussed in Section \ref{sec:inclination}.

\section{Discussion and Conclusions}\label{sec:discussion}

One of our most striking results is that for viewing angles $\le 50^\circ$,  rotating EVPAs manifested as loops in the $\mathcal{Q}-\mathcal{U}$ plane are copiously produced by GRMHD simulations in the MAD regime.  
Loops are largely associated with violent flux eruptions leading to enhanced emissivity of disk material that is compressed against the orbiting flux bundles. 

The loops in linear polarization also have a clear signature in the cross-correlation of $\mathcal{Q}$ and $\mathcal{U}$ as the correlation is maximized by phase shifting one of the observables by a quarter of a period. This allows us to deduce a typical loop period between $27$ and $53$ minutes.  There is a tendency for smaller phase shifts (ie. faster periodicity) in co-rotating cases compared to counter-rotating cases which imply that the loop periods are sensitive to black hole spin.  While the cross-correlation peaks in our data are fairly broad which makes it difficult to determine exact phase shifts, the clear trends we obtain suggest that polarization time-domain data can provide an interesting avenue to constrain orbital motion and ultimately black hole spin.  

Another striking feature shows up in the lightcurves of different Stokes parameters: while total intensity displays typical red-noise accretion variability, we observe a broad but significant variability excess in \qstokes~ and \ustokes~ at periods of $\approx 1\rm\, hour$. This polarized excess variability can be explained as follows: while all Stokes vectors vary according to the fluctuations imprinted by the mass accretion rate (corresponding to radial motions), $\mathcal{Q}$ and $\mathcal{U}$ are additionally sensitive to the azimuthal motions of emitting features such that they sample the orbital dynamics of the innermost accretion flow.  
It will be very interesting to search for similar time-domain signatures in observed linear polarization lightcurves.  

In our sample, polarization loops are also visible for SANE cases with disk-dominated emission ($R_{\text{high}}=1$). In this case they are produced by low-$m$ spiral modes in the turbulent accretion flow itself.  
The shot noise due to multiple uncorrelated regions across the disk leads to a lower polarization overall compared to MAD loops (and thus loops with smaller radii in the $\mathcal{Q}-\mathcal{U}$ plane), however, there is sufficient power in $m=1$ and $m=2$ modes so that net rotating EVPAs are recovered.  
That being said, it is important to stress that both from theoretical grounds which demand $T_e<T_{\rm ions}$ and from observational modeling of Sgr A* \citep{EventHorizonTelescopeCollaborationAkiyamaEtAl2022a}, these models (SANE, $R_{\text{high}}=1$) are disfavored. Already at $R_\text{high}=10$ we do not recover any polarization loops with hourly periods any more. This suggests that the observed loops by \cite{wielgus22} are generated by a accretion flow that is in the MAD regime.

MAD flux eruptions strongly perturb the magnetic field in the emitting region, resulting in local polarization vectors which differ significantly from an average ``background'' magnetic field/EVPA configuration. Emitting spiral features in SANE simulations on the other hand simply enhance the local polarized emissivity, while keeping the direction of the EVPA largely intact.  This means that inferring magnetic field topology from ``hotspot'' models should be done with some care since (MAD-) flares are capable of significantly perturbing the magnetic field in the emitting region.  
For the more docile $R_\text{high}=1$ SANE simulations however, our results lend some support to treating the magnetic field as a fixed background as commonly done in hotspot models \citep[e.g.][]{VosMoscibrodzkaEtAl2022}.  This in principle allows to relate the orbital frequency of emitting features with the period in the $\mathcal{Q}-\mathcal{U}$ plane: in the SANE case which is dominated by the toroidal magnetic field component, as the background EVPA rotates twice along one orbit (cf. Figure \ref{fig:multi_sane_xl}), the typical orbital period corresponds to twice the polarization period (see also the discussion in \cite{GravityCollaborationAbuterEtAl2018,VosMoscibrodzkaEtAl2022}).  

Regardless of the inclination, in our study the polarization loops always enclose the $\mathcal{P}=0$ origin whereas the ALMA data reported in \cite{wielgus22} features a clear offset with $\overline{\mathcal{P}} =0.19 \rm~ Jy$. In the latter study, the authors argue that the total polarisation flux is a sum of the two components: $\mathcal{P} = \mathcal{P}_{\rm hsp} + \mathcal{P}_{\rm sh}$, where hsp stands for the varying hotspot and sh stands for the background black hole shadow emission with $\mathcal{P}_{\rm sh} \approx 0.2 \rm~ Jy $. 
 Since our model includes the entire emitting region within $0.2\rm~mas$, any background shadow emission is consistently taken into account.  In addition, the mean linear polarization degree in all our MAD models is relatively low, in particular below $5-10\%$ fractional linear polarization commonly observed from Sgr A*. While the obtained polarization fractions are quite sensitive to the uncertain electron thermodynamics, the lack of an overall offset in our simulations could be an indication that another polarized emission component outside our field of view might be required.  

Despite the different underlying assumptions between the GRMHD simulations and the hotspot models of \citep{wielgus22,VosMoscibrodzkaEtAl2022} similar features can be produced by both models. Specifically, our $t_3$ loop shown in Figure~\ref{fig:loops-lightcurves-mad} is very similar to the pretzel-looking loop shown in Figure 6 of \cite{VosMoscibrodzkaEtAl2022}. The period of both loops is about $180\rm~M\, \sim 60~\rm min$ and similar periods are reported by GRAVITY for a number of flares \citep{TheGRAVITYCollaborationAbuterEtAl2023}. 
 
The features observed so far (GRAVITY and ALMA loops) move in a clockwise direction whereas loops and emitting region in our simulations rotates counter-clockwise. We have verified that clockwise rotation is obtained when we choose an inclination $>90^\circ$; this data is shown in Appendix \ref{sec:flipped}. Apart from the directionality, our conclusions remain unaffected when flipping the viewing angle.  

There are several recent studies that investigate GRMHD simulations in the MAD regime to model black hole flares \citep[e.g.][]{Chatt21,ScepiDexter22,RipperdaLiskaEtAl2022,JiaRipperdaEtAl2023}.  Although we focus on the polarization properties, it is interesting to also compare the behaviour of total intensity during the flares.  In particular, \cite{JiaRipperdaEtAl2023} found that except for very cold disk electrons $(R_{\rm low},R_{\rm high})=(100,100)$, the 230\,GHz flux dims during eruptions.  We instead find that sharp flux increases during the eruption are recovered, in particular when the flux is above $\approx 2~\rm Jy$ (cf. Figure~\ref{fig:loops-lightcurves-mad}).  
The difference is likely explained by the larger optical depth obtained in our simulations which have been scaled to Sgr A*, not M87* as in the comparison paper.  Indeed, as \cite{JiaRipperdaEtAl2023} discussed, the dimming can be understood from the temperature dependence of the emissivity alone.  This explanation, however, only holds in the optically thin regime.  In Sgr A* models, the optical depth is close to one in quiescence and can reach $\tau=10$ at the eruption site.  Since we model thermal emission and the Planck function increases monotonously with temperature for any frequency, flux increase is instead expected during optically thick flaring events.    

On a related note, the $230~\rm GHz$ spectral indices obtained in our fiducial model ($\alpha\in[-1,-0.2]$) are too small to match the observations which find $\alpha\ge0$ in quiescence and only slightly negative $\approx -0.2$ post flare \citep{Goddi2021,WielgusMarchiliEtAl2022}.  
This implies that our models might still be too optically thin.  In fact, as also shown by \cite{RicarteGammieEtAl2023}, for the adopted electron temperature prescriptions following Eq.~(\ref{eq:rhigh}) with $R_{\rm high}\in\{1,10,40,160\}$ the spectral indices are negative for essentially all the SANE and MAD models considered here.  
We suggest that models with colder electrons should be tried for Sgr A* which will increase the plasma density and optical depth of the models and thereby increase the spectral index.  

Since EHT and GRAVITY should be able to resolve the centroid of the flaring emission on a scale of $\sim 10 \rm~M$, it will be interesting to assess the imprint of the flux eruptions on interferometric observations.  In particular, it is important to predict the radius and motion of the emission centroid. We leave such an investigation to future work.  
Furthermore, including radiative cooling in the GRMHD simulations might be worthwhile since fast-cooling heated electrons can cool back into the mm-emitting range on a short timescale, thereby changing also the mm-lightcurves.  

\section*{Acknowledgements}

We would like to acknowledge Hector Olivares and Christian Fromm for stimulating discussions during this work.  We thank Maciek Wielgus for useful comments on a draft.  
JD is supported by NASA grant NNX17AL82G and a Joint Columbia/Flatiron Postdoctoral Fellowship. Research at the Flatiron Institute is supported by the Simons Foundation. 
YM is supported by the National Natural Science Foundation of China (grant no. 12273022) and the Shanghai Municipality orientation program of basic research for international scientists (grant no. 22JC1410600).
This research has made use of NASA's Astrophysics Data System.

{\it Software:} {\tt python} \citep{travis2007,jarrod2011}, {\tt scipy} \citep{jones2001}, {\tt numpy} \citep{walt2011}, and {\tt matplotlib} \citep{hunter2007}.

\section*{Data Availability}
The data underlying this article will be shared on reasonable request to the corresponding author.

\bibliographystyle{mnras}
\bibliography{references.bib}

\appendix

\section{Sample of loops for the eruption and re-accretion phases}\label{sec:all-loops}
\begin{figure}
    \centering \includegraphics[width=0.5\textwidth]{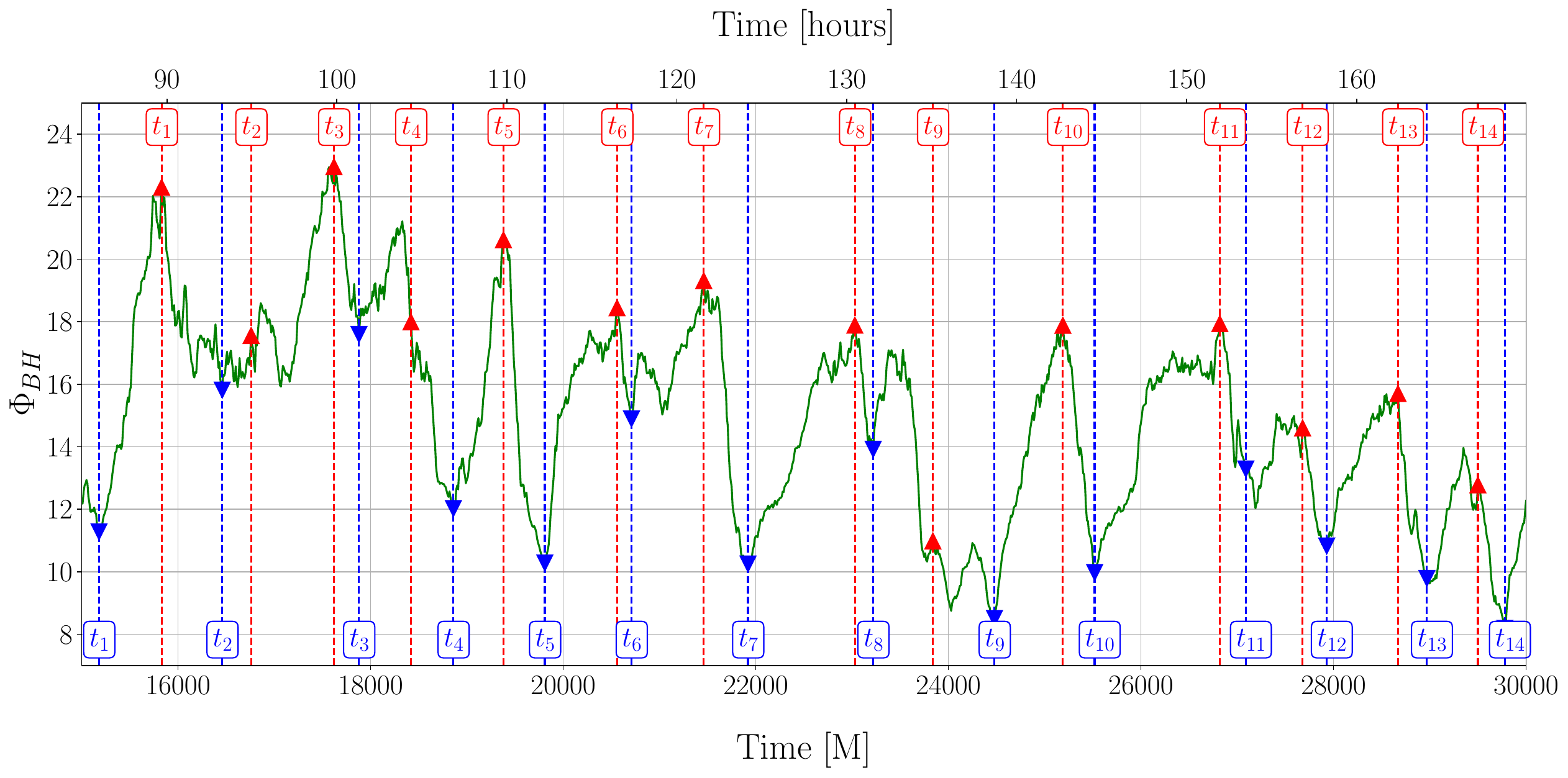}
    \caption{The sample along maxima and minima of the horizon penetrating flux for the fiducial case with $a_*=-0.9375$, $R_{\rm high}=1$, $i=10^\circ$.}
    \label{fig:phib-minmax}
\end{figure}

\begin{figure*}
    \centering    \includegraphics[width=0.8\textwidth]{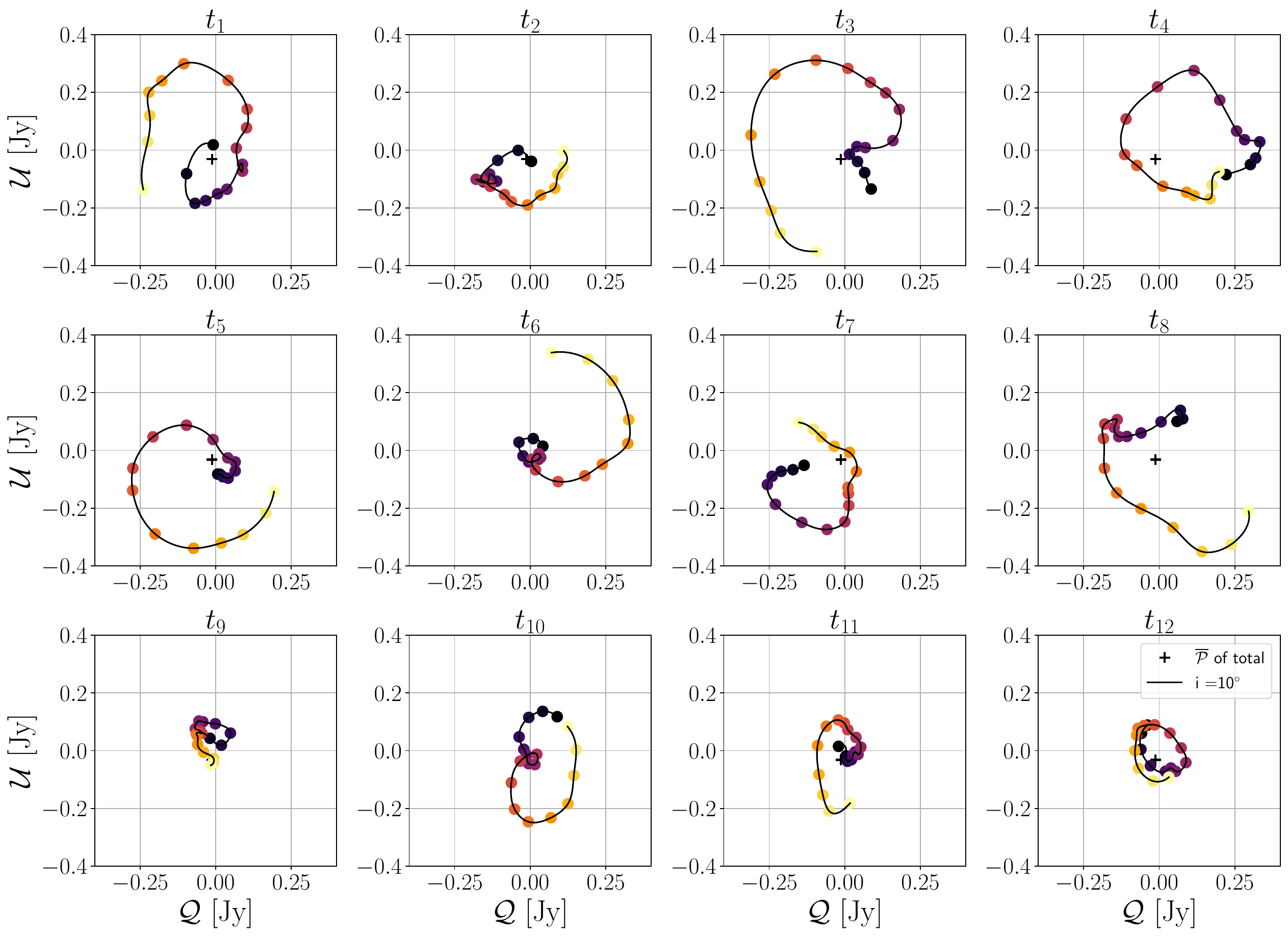}
    \caption{The loops corresponding to the local maxima of the magnetic flux. (MAD, $R_{\rm high}$ = 1, $a_*=-0.9375$)}
    \label{fig:loops-max}

    \centering
\includegraphics[width=0.8\textwidth]{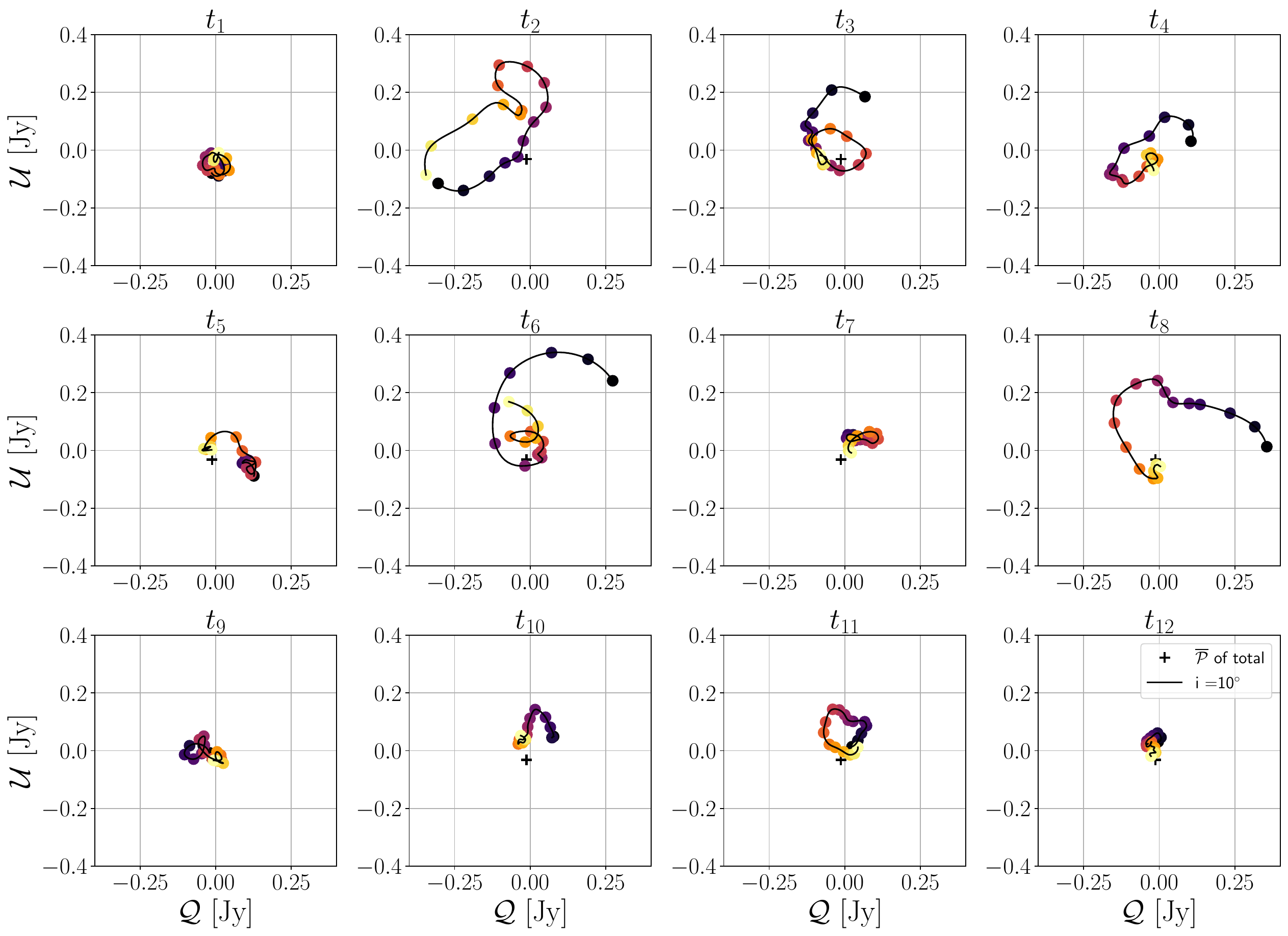}
    \caption{The loops corresponding to the local minima of the magnetic flux. (MAD, $R_{\rm high} = 1$, $a_*=-0.9375$)}
    \label{fig:loops-min}

\end{figure*}

The fiducial inclination for our analysis is $i=10^{\circ}$, which reveals loops that rotate counter-clockwise. In Figure \ref{fig:flipped_inc} we show that flipping the inclination to $i=170^{\circ}$ still generates loops that rotate clockwise and are slightly different in shape compared to their counterparts in Figure \ref{fig:loops-lightcurves-mad}.
The majority of the loops in the $\mathcal{Q-U}$ plane form during the eruptive period of the magnetic flux. In order to be able to inspect this correlation in more detail, we show the $\mathcal{Q-U}$ plane during a selection of the eruptive and accumulating periods of the magnetic flux. For an overview of the sample, see Figure~\ref{fig:phib-minmax}, 

the maxima of the magnetic flux are marked in red triangles and the minima in blue. These time markers are the initial time of the inspection when we begin to monitor the $\mathcal{Q-U}$ as the magnetic flux evolves for about 1 hour. In Figure~\ref{fig:loops-max}. we see the corresponding $\mathcal{Q-U}$ for the selection of eruptive periods; in almost all panels, we can recognize a loopy feature. On the other hand, for the re-accretion periods in Figure~\ref{fig:loops-min}, most panels depict a more stochastic pattern than the loop-shaped curves. 

\section{$R_\text{\lowercase{high}}$ dependence of SANE loops}\label{sec:Rhigh}

\begin{figure}
    \centering \includegraphics[width=0.4\textwidth]{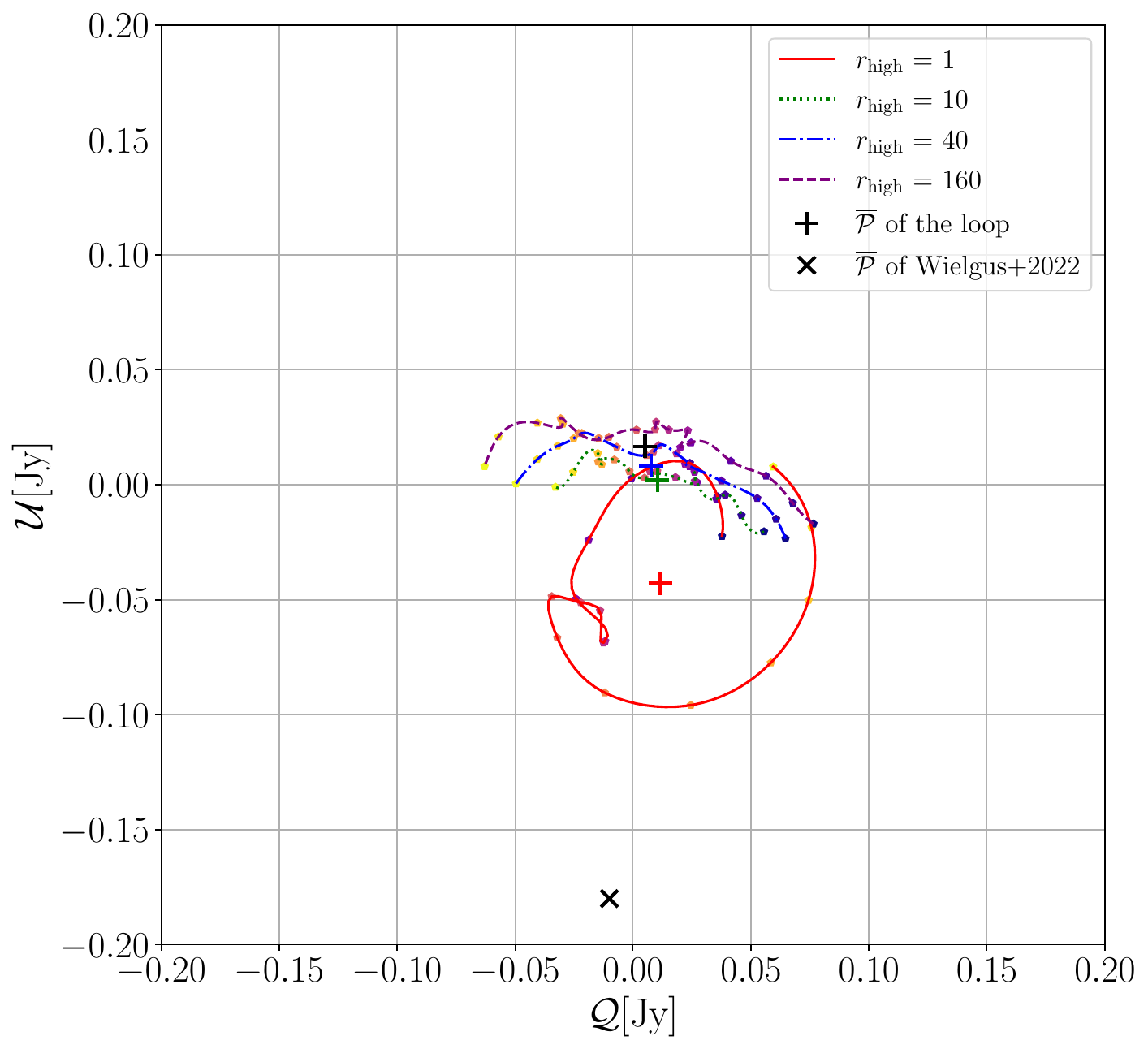}
    \caption{The effect of $R_\text{high}$ on the properties of the $\mathcal{Q-U}$ plane in SANE model. Only the $R_\text{\lowercase{high}}=1$ model can produce loops ($a_*=-0.9375$, $i=10^\circ$). See \url{https://doi.org/10.5281/zenodo.8302230} for an animation of the $R_\text{high}=1$ and $R_\text{high}=160$ cases.}
    \label{fig:Rhighs}
\end{figure}

The response of the polarization signal upon changing the electron temperature parameter $R_\text{high}$ is shown in Figure \ref{fig:Rhighs} for a fiducial SANE simulation.  We choose an interval of an hour starting at $25260~\rm M$, thus within the time-span also shown in Figure \ref{fig:multi_sane_xl}.  
Hourly polarization loops are only recovered with our most disk-dominated model $R_\text{high}=1$.  For larger values, motion in the \qstokes-\ustokes~ plane is slower with a stronger stochastic component.  There are two main underlying effects behind this: 1. the change of the emission region which moves to larger distances hence slower timescales, 2. the increase of the Faraday depth when cooling of the electrons in the disk contributing to more stochastic polarization signals. 

\section{Flipped viewing angle}\label{sec:flipped}

\begin{figure}
    \centering \includegraphics[width=0.6\textwidth]{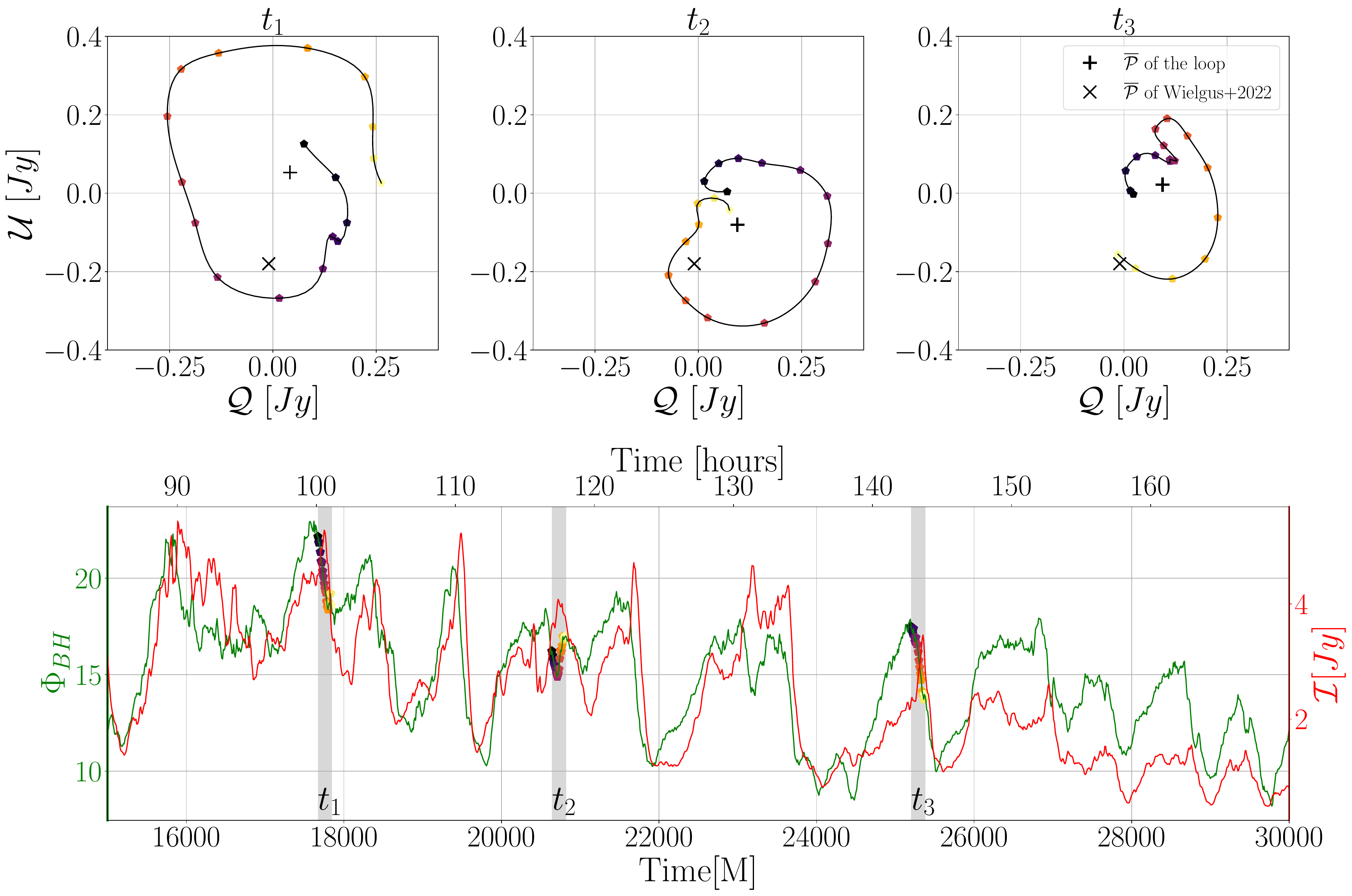}
    \caption{Flipped inclination of the selected loops in Figure \ref{fig:loops-lightcurves-mad} (MAD, $a_*=-0.9375$, $R_{\rm high}=1$, $i=170^\circ$).}
    \label{fig:flipped_inc}
\end{figure}

Here we show the selected flux eruption events from Figure \ref{fig:loops-lightcurves-mad} when the viewing angle is flipped with respect to the equatorial plane.  Figure \ref{fig:flipped_inc} thus adopts a viewing angle of $170^\circ$, showing that the loops now move clockwise in the $\mathcal{Q-U}$ plane.  Since the track is sensitive to small scale asymmetries in the accretion flow, the loops are not exactly mirrored versions of Figure \ref{fig:loops-lightcurves-mad}, however their overall extent and period remains similar.  

\bsp	
\label{lastpage}
\newpage

\end{document}